\def\be{\begin{equation}}
\def\ee{\end{equation}}
\def\ba{\begin{array}}
\def\ea{\end{array}}
\def\1{{\bf 1}}
\def\p{\prime}
\def\pp{{\prime\prime}}
\def\t{\tau}
\def\R1{{1\!\! 1}}
\def\Rb{{I\!\! R}}
\def\Nb{{I\!\! N}}
\def\Fb{{I\!\! F}}
\def\Cb{\ \hbox{\vrule width 0.6pt height 6pt depth 0pt
		      \hskip -3.5 pt} C}
\documentstyle[12pt]{article}
\begin{document}
\parskip=3pt
\parindent=18pt
\baselineskip=20pt
\setcounter{page}{1}
\centerline{\Large\bf Symmetry, Integrable Chain Models}
\vspace{2ex}
\centerline{\Large\bf and Stochastic Processes}
\vspace{6ex}
\centerline{\large{\sf Sergio Albeverio$^\star$} ~~~and~~~ {\sf Shao-Ming Fei}
\footnote{\sf Alexander von Humboldt-Stiftung fellow.\\
\hspace{5mm}On leave from Institute of Physics, Chinese Academy of Sciences,
Beijing}}
\vspace{4ex}
\parindent=40pt
{\sf Institute of Mathematics, Ruhr-University Bochum,
D-44780 Bochum, Germany}\par
\parindent=35pt
{\sf $^\star$SFB 237 (Essen-Bochum-D\"usseldorf); 
BiBoS (Bielefeld-Bochum);\par
\parindent=40pt
CERFIM Locarno (Switzerland)}\par
\vspace{6.5ex}
\parindent=18pt
\parskip=5pt
\begin{center}
\begin{minipage}{5in}
\vspace{3ex}
\centerline{\large Abstract}
\vspace{4ex}
A general way to construct chain models with certain
Lie algebraic or quantum Lie algebraic symmetries is presented.
These symmetric models give rise to series of integrable systems. 
As an example the chain models with $A_n$ symmetry and the related
Temperley-Lieb algebraic structures and representations are discussed. It
is shown that corresponding to these $A_n$ symmetric integrable chain models
there are exactly solvable stationary discrete-time (resp.
continuous-time) Markov chains whose spectra of the
transition matrices (resp. intensity matrices)
are the same as the ones of the corresponding integrable models.
\end{minipage}
\end{center}
\newpage
\tableofcontents
\section{Introduction}

Integrable chain models have been discussed for many years
in statistical and condensed matter physics. Some of them
have been obtained and investigated using an algebraic ``Bethe Ansatz
method" \cite{faddeev}, see e.g., \cite{2345}
for periodic boundary conditions and \cite{67} for fixed boundary
conditions. The intrinsic symmetry of these integrable chain models 
plays an essential role in finding complete sets of
eigenstates of the systems.

On the other hand, stochastic models like 
stochastic reaction-diffusion models,
models describing coagulation/decoagulation, 
birth/death processes, pair-creation and 
pair-annihilation of molecules on a chain, 
have attracted considerable interest due to their importance in many physical,
chemical and biological processes \cite{evan}. Some
of these stochastic models can be ``exactly solved", see e.g., \cite{prl}.
The theoretical description of stochastic reaction-diffusion systems
is given by the ``master equation'' which describes the time evolution of the 
probability distribution function \cite{4s}. This equation has the form of 
a heat equation with potential (i.e.,
a Schr\"odinger equation with ``imaginary time"). If an integrable system can 
be transformed into a stochastic reaction-diffusion system, e.g., by a unitary
transformation between their respective Hamiltonians, looked upon
as self-adjoint operators acting in the respective Hilbert spaces,
then the stochastic 
model so obtained is exactly soluble with the same energy spectrum as the 
one of the integrable system \cite{henkel}.

In this paper, by using the coproduct properties of bi-algebras, we 
present a general procedure for the construction of chain models having
a certain Lie algebra or
quantum Lie algebra symmetry with nearest
or non-nearest neighbours interactions. The models obtained in this way can
be reduced to integrable ones via a detailed representation of the
symmetry algebras involved. As an example we discuss integrable models with
$A_n$ symmetry. These models turn out to have an additional Temperley-Lieb
(TL) algebraic structure, in the sense that the Hamiltonians give rise
to unitary 
representations of the TL algebra and can be expressed by the elements of 
the TL algebra. What is more, we find all these models can be transformed 
into both stationary discrete-time and stationary continuous-time Markov
chains (discrete reaction-diffusion models, see e.g.\cite{678}),
whose spectra of the transition matrices resp. intensity matrices
are the same as the ones of these $A_n$ invariant integrable models.

In section 2 we first recall the basic properties of bialgebras
and then describe their use in the construction of
chain models with a Lie algebra symmetry resp. a quantum Lie algebra
symmetry. In section 3 we discuss the $A_n$ symmetric integrable models 
in the fundamental representation of $A_n$. We give the representation
of a TL algebra in these models. In section 4 we prove that these $A_n$ 
symmetric integrable chain models can be transformed into both
continuous-time and discrete-time Markov chains.
Some conclusions and remarks are given in section 5.

\section{Chain Models with Algebraic Symmetry}
\subsection{Bi-algebra}
Let $A$ be an associative algebra. $A$ is said to be a bi-algebra if it
contains two linear operators, the multiplication $m$ and the coproduct
$\Delta$.

The operation of multiplication $m$ is defined by:

\be\label{1}
\ba{l}
m:~A\otimes A\to A\\[3mm]
m(a\otimes b)=ab,~~\forall a,b\in A.
\ea
\ee
Let $\{e_i\}$ be the set of base elements of the algebra $A$. Then (\ref{1})
means
\be\label{2}
m(e_i\otimes e_j)=\sum_{k}m_{ij}^k e_k.
\ee
The tensor $m_{ij}^k$ determines the properties of $m$ completely.

The multiplication $m$ is associative,
\be\label{3}
m(m\otimes {\bf id})=m({\bf id}\otimes m),
\ee
i.e.,
\be\label{4}
\sum_k m_{ij}^k m_{kl}^n=\sum_k m_{ik}^n m_{jl}^k,
\ee
where ${\bf id}$ denotes the ~identity transformation,
$$
{\bf id}:~A\to A,~~~~~~{\bf id}(a)=a.
$$
However in general $m$ is not commutative, i.e., we have
\be\label{5}
m\circ p\neq m,
\ee
or equivalently, $m_{ij}^k\neq m_{ji}^k$, 
where $p$ is the transposition operator,
\be\label{6}
p:~A\otimes A\to A\otimes A,~~p(a\otimes b)=(b\otimes a),~~\forall
a,b\in A.
\ee

The coproduct operator $\Delta$ maps $A$ into $A\otimes A$:
\be\label{7}
\ba{l}
\Delta:~A\to A\otimes A,\\[3mm]
\Delta(e_i)=\displaystyle\sum_{jk} \mu^{jk}_i e_j\otimes e_k.
\ea
\ee
The properties of the coproduct are described by the tensor $\mu^{jk}_i$.

$\Delta$ is an algebraic homomorphism,
\be\label{8}
\ba{l}
\Delta(ab)=\Delta(a)\Delta(b),~\forall a,b\in A,\\[3mm]
\displaystyle\sum_{k}\mu^{rs}_k m_{ij}^k=
\displaystyle\sum_{ktpq} m_{kp}^r
m_{tq}^s\mu^{kt}_i\mu^{pq}_j,
\ea
\ee
where $\Delta(a)$ and $\Delta(b)$ belong to $A\otimes A$ and the
multiplication of tensors is defined by
$$
(a_1\otimes a_2)(b_1\otimes b_2)=a_1b_1\otimes
a_2b_2,~~~~a_1,a_2,b_1,b_2\in A.
$$
The coproduct is associative
\be\label{9}
\ba{l}
(\Delta\otimes{\bf id})\Delta=({\bf id}\otimes\Delta)\Delta,\\[3mm]
\displaystyle\sum_{t}\mu^{rs}_t\mu^{tp}_i=
\displaystyle\sum_{t}\mu^{sp}_t\mu^{rt}_i,
\ea
\ee
but in general not co-commutative
\be\label{10}
p\circ\Delta\neq\Delta,~~~~
\mu^{rs}_j\neq\mu^{sr}_j.
\ee
The operation $\Delta$ preserves all the algebraic relations of the algebra
$A$. It gives a way to find representations of the algebra $A$ in the
direct product of spaces.

If a bi-algebra has in addition unit, counit and antipode
operators, then it is called a Hopf algebra. Lie algebras are Hopf
algebras with $\Delta$ co-commutative. Quantum algebras are Hopf algebras
that are not co-commutative, see e.g. \cite{pressley} and references 
therein.

\subsection{Chain Models with Lie algebraic Symmetry}
Let $A$ be a Lie algebra with basis $e=\{e_\alpha\}$,
$\alpha=1,2,...,n$, satisfying the Lie commutation relations
\be\label{11}
[e_\alpha,e_\beta]=C_{\alpha\beta}^\gamma e_\gamma,
\ee
where $C_{\alpha\beta}^\gamma$ are the structure constants with respect
to the base $e$.

Let $\Delta$ (resp. $C(e)$) be the coproduct operator (resp. Casimir
operator) of the algebra $A$. We have
\be\label{12}
[C(e),e_\alpha]=0,~~~\alpha=1,2,...,n.
\ee
The coproduct operator action on the Lie algebra elements is given by
\be\label{13}
\Delta e_\alpha=e_\alpha\otimes\1+\1\otimes e_\alpha,
\ee
$\1$ stands for the identity operator. It is easy to check that
$$
[\Delta e_\alpha,\Delta e_\beta]=C_{\alpha\beta}^\gamma \Delta e_\gamma.
$$
From the properties of the coproduct, $\Delta C(e)$ is a rank two tensor
satisfying
\be\label{14}
[\Delta C(e),\Delta e_\alpha]=0,~~~\alpha=1,2,...,n.
\ee

Let $\Fb$ denote an entire function defined on the $(L+1)$-th tensor space 
$A\otimes A\otimes...\otimes A$ of the algebra $A$. From (\ref{14}) we have
\be\label{15}
[\Fb(\Delta C(e)),\Delta e_\alpha]=0,~~~\alpha=1,2,...,n.
\ee

We consider ``a chain with $L+1$ sites", i.e., the set $\{1,2,...,L+1\}$.
We call it an algebraic chain in the following.
To each point $i$ of the chain we associate a (finite ~dimensional
complex) Hilbert space $H_i$. We can then associate to the whole chain
the tensor product $H_1\otimes H_2\otimes...\otimes H_{L+1}$.
For simplicity we use subindices $i,j,k...$ for the points in the chain
sites.

The generators of the algebra $A$ acting on the Hilbert space
$H_1\otimes H_2\otimes...\otimes H_{L+1}$ associated with the above chain
are given by
\be\label{16}
E_\alpha=\Delta^L e_\alpha,~~~\alpha=1,2,...,n,
\ee
where we have defined
\be
\label{17}
\Delta^m=(\underbrace{{\bf id}\otimes ... \otimes{\bf id}}_{m~times}\otimes\Delta)
...({\bf id}\otimes{\bf id}\otimes\Delta)
({\bf id}\otimes\Delta)\Delta,~~\forall m\in\Nb.
\ee
$E_\alpha$ also generates the Lie algebra $A$,
$$
[E_\alpha,E_\beta]=C_{\alpha\beta}^\gamma E_\gamma.
$$

We call 
\be\label{18}
H=\sum_{i=1}^L \Fb(\Delta C(e))_{i,i+1}
\ee
the (quantum mechanics) Hamiltonian associated with the chain $\{1,2,...,L+1\}$
and given by real entire function $\Fb$. Here $\Fb(\Delta C(e))_{i,i+1}$
means that the rank-two tensor
element $\Fb(\Delta C(e))$ is on the sites $i$ and $i+1$ of the
chain, i.e.,
$$
\Fb (\Delta C(e))_{i,i+1}=\1_1\otimes...\otimes\1_{i-1}\otimes \Fb(\Delta
C(e))\otimes\1_{i+2}\otimes... \otimes\1_{L+1}.
$$

{\sf [Theorem 1]}. The Hamiltonian $H$ is a self-adjoint operator acting
in $H_1\otimes H_2\otimes...\otimes H_{L+1}$ and is invariant under the
algebra $A$.
	
{[\sf Proof].} That $H$ is self-adjoint is immediate from the definition.
To prove the invariance of $H$
it suffices to prove $[H,E_\alpha]=0$, $\alpha=1,2,...,n$.
From formula (\ref{13}) $E_\alpha$ in (\ref{16}) is simply
\be\label{19}
E_\alpha=\sum_{i=1}^L(e_{\alpha})_i,
\ee
where
$$
(e_{\alpha})_i=\1_1\otimes...\otimes\1_{i-1}\otimes 
(e_{\alpha})\otimes\1_{i+1}\otimes... \otimes\1_{L+1}.
$$
Obviously
\be\label{20}
[ \Fb(\Delta C(e))_{i,i+1},(e_{\alpha})_j]=0,~~~\forall j\neq i,i+1.
\ee
By using formula (\ref{20}) and (\ref{13}) we have
$$
\ba{rcl}
[H,E_\alpha]&=&
\left[\displaystyle\sum_{i=1}^L \Fb(\Delta C(e))_{i,i+1},
\displaystyle\sum_{j=1}^L(e_{\alpha})_j\right]\\[5mm]
&=&\displaystyle\sum_{i=1}^L\left[\Fb(\Delta C(e))_{i,i+1},
\displaystyle\sum_{j=1}^{i-1}(e_{\alpha})_j+
\displaystyle\sum_{j=i+2}^{L}(e_{\alpha})_j
+(e_{\alpha})_i+(e_{\alpha})_{i+1}\right]\\[5mm]
&=&\displaystyle\sum_{i=1}^L\left[\Fb(\Delta C(e))_{i,i+1},
(e_{\alpha})_i+(e_{\alpha})_{i+1}\right]\\[5mm]
&=&\displaystyle\sum_{i=1}^L\left[
\Fb(\Delta C(e))_{i,i+1},(\Delta e_\alpha)_{i,i+1}\right].
\ea
$$
By formula (\ref{15}) we get $[H,E_\alpha]=0$, $\alpha=1,2,...,n$.
\hfill $\rule{3mm}{3mm}$

The Hamiltonian system given by (\ref{18}) describes
nearest neighbours interactions. Systems with $A$-invariant
Hamiltonians and non-nearest neighbours interactions can be constructed by
iterating the application of the coproduct operator to the Casimir 
operator. For instance, the following Hamiltonian describes a system with
$N+1(N\leq L)$ sites interactions:
\be\label{21}
H^N=\sum_{i=1}^{L-N+1}\Fb(\Delta^N C(e))_{i,i+1,...,i+N},
\ee
with $\Delta^N$ as in definition (\ref{17}).
$H^N$ commutes with the generators $E_\alpha$, $\alpha=1,2,...,n$,
of the algebra $A$ on the Hilbert space $H_1\otimes ...\otimes H_{L+1}$
since
$$
\ba{rcl}
[H^N,E_\alpha]&=&
\left[\displaystyle\sum_{i=1}^{L-N+1} \Fb(\Delta^N C(e))_{i,i+1,...,i+N},
\displaystyle\sum_{j=1}^L(e_{\alpha})_j\right]\\[5mm]
&=&\displaystyle\sum_{i=1}^{L-N+1}\left[\Fb(\Delta^N C(e))_{i,i+1,...,i+N},
\displaystyle\sum_{j=1}^{i-1}(e_{\alpha})_j
+\displaystyle\sum_{j=i}^{i+N}(e_{\alpha})_j
+\displaystyle\sum_{j=i+N+1}^{L}(e_{\alpha})_j\right]\\[5mm]
&=&\displaystyle\sum_{i=1}^{L-N+1}\left[\Fb(\Delta^N C(e))_{i,i+1,...,i+N},
\displaystyle\sum_{j=i}^{i+N}(e_{\alpha})_j\right]\\[5mm]
&=&\displaystyle\sum_{i=1}^{L-N+1}\left[\Fb(\Delta^N C(e)),
(\Delta^N e_\alpha)\right]_{i,i+1,...,i+N}=0,
\ea
$$
where the relation $\Delta^N ([\Fb(C(e)),e_\alpha])
=[\Delta^N(\Fb(C(e))),\Delta^N(e_\alpha)]=0$ has been used.

\subsection{Chain Models with Quantum Lie Algebraic Symmetry}

Let $e=\{e_\alpha,f_\alpha,h_\alpha\}$, $\alpha=1,2,...,n$, 
be the Chevalley basis of a Lie algebra $A$ with rank $n$.
Let $e^\p=\{e_\alpha^\p,f_\alpha^\p,h_\alpha^\p\}$, 
$\alpha=1,2,...,n$, be the corresponding elements of the
quantum (q-deformed) Lie algebra  $A_q$. We denote by $r_\alpha$ the simple
roots of the Lie algebra $A$. The Cartan matrix $(a_{\alpha\beta})$ is then
\be\label{22}
a_{\alpha\beta}=\frac{1}{d_\alpha}(r_\alpha\cdot r_\beta),~~~
d_\alpha=\frac{1}{2}(r_\alpha\cdot r_\alpha).
\ee

We introduce a complex quantum parameter $q$, such that 
$q^{d_\alpha}\neq\pm1,0$. The
quantum algebra generated by $\{e_\alpha^\p,f_\alpha^\p,h_\alpha^\p\}$
is defined by the following relations:
\be\label{23}
\ba{rcl}
[h_\alpha^\p,h_\beta^\p]&=&0,\\[4mm]
[h_\alpha^\p,e_\beta^\p]&=&a_{\alpha\beta}e_\beta^\p,\\[4mm]
[h_\alpha^\p,f_\beta^\p]&=&-a_{\alpha\beta}f_\beta^\p,\\[4mm]
[e_\alpha^\p,f_\beta^\p]&=&\delta_{\alpha,\beta}
\displaystyle\frac{q^{d_\alpha h_\alpha^\p}-q^{-d_\alpha h_\alpha^\p}}
{q^{d_\alpha}-q^{-d_\alpha}}
\ea
\ee
together with the quantum Serre relations
\be\label{24}
\ba{l}
\displaystyle\sum_{\gamma=0}^{1-a_{\alpha\beta}}(-1)^\gamma
\left[\ba{c}1-a_{\alpha\beta}\\[1mm]\gamma\ea\right]_{q^{d_\alpha}}
(e_\alpha^\p)^\gamma e_\beta^\p(e_\alpha^\p)^{1-a_{\alpha\beta}-
\gamma}=0,~~~i\neq j, \\[6mm]
\displaystyle\sum_{\gamma=0}^{1-a_{\alpha\beta}}(-1)^\gamma
\left[\ba{c}1-a_{\alpha\beta}\\[1mm]\gamma\ea\right]_{q^{d_\alpha}}
(f_\alpha^\p)^\gamma f_\beta^\p(f_\alpha^\p)^{1-a_{\alpha\beta}-
\gamma}=0,~~~i\neq j,
\ea
\ee
where for $m\geq n\in\Nb$,
$$
\ba{rcl}
\left[\ba{c}m\\[1mm]n\ea\right]_{q}&=&\displaystyle\frac{[m]_q!}
{[n]_q![m-n]_q!},\\[7mm]
[n]_q!&=&[n]_q[n-1]_q...[2]_q[1]_q,\\[5mm]
[n]_q&=&\displaystyle\frac{q^n-q^{-n}}{q-q^{-1}}.
\ea
$$

The coproduct operator $\Delta^\prime$ of the quantum algebra $A_q$ is
given by
\begin{eqnarray}
\Delta^\p h_\alpha^\p&=&h_\alpha^\p\otimes\1+\1\otimes h_\alpha^\p,\label{25}\\[3mm]
\Delta^\p e_\alpha^\p&=&e_\alpha^\p\otimes q^{-d_\alpha h_\alpha^\p}
+q^{d_\alpha h_\alpha^\p}\otimes e_\alpha^\p,\label{26}\\[3mm]
\Delta^\p f_\alpha^\p&=&f_\alpha^\p\otimes q^{-d_\alpha h_\alpha^\p}
+q^{d_\alpha h_\alpha^\p}\otimes f_\alpha^\p.\label{27}
\end{eqnarray}
It is straightforward to check that $\Delta^\p$ preserves all the algebraic
relations in (\ref{23}) and (\ref{24}).

Let $C_q(e^\prime)$ be the Casimir operator of $A_q$, i.e.,
$[C_q(e^\prime), a]=0,~\forall a\in A_q$. For any entire function $\Fb$ 
of $C_q(e^\prime)$, we have
\be\label{28}
[\Fb(C_q(e^\prime)), a]=0,~\forall a\in A_q
\ee
and
\be\label{29}
[\Delta^\p \Fb(C_q(e^\prime)), \Delta^\p a]=0,~\forall a\in A_q.
\ee
Especially, by formula (\ref{25}) one gets
\be\label{30}
\Delta^\p q^{\pm d_\alpha h_\alpha^\p}=q^{\pm d_\alpha h_\alpha^\p}\otimes 
q^{\pm d_\alpha h_\alpha^\p}.
\ee
Hence
\be\label{31}
[\Delta^\p \Fb(C_q(e^\prime)), \Delta^\p q^{\pm d_\alpha h_\alpha^\p}]
=[\Delta^\p \Fb(C_q(e^\prime)), q^{\pm d_\alpha h_\alpha^\p}\otimes 
q^{\pm d_\alpha h_\alpha^\p}]=0.
\ee

The generators of $A_q$ on a chain with $(L+1)$-sites are given by
\be\label{32}
\ba{rcl}
H^\p_\alpha&=&\Delta^{\p L} h^\p_\alpha=
\displaystyle\sum_{i=1}^{L+1}(h^\p_\alpha)_i\\[4mm]
E^\p_\alpha&=&\Delta^{\p L} e^\p_\alpha,\\[4mm]
&=&\displaystyle\sum_{i=1}^{L+1}
(q^{d_\alpha h^\p_\alpha})_1\otimes...
\otimes (q^{d_\alpha h^\p_\alpha})_{i-1}\otimes(e^\p_\alpha)_i\otimes
(q^{-d_\alpha h^\p_\alpha})_{i+1}\otimes...
\otimes(q^{-d_\alpha h^\p_\alpha})_{L+1},\\[6mm]
F^\p_\alpha&=&\Delta^{\p L} f^\p_\alpha\\[4mm]
&=&\displaystyle\sum_{i=1}^{L+1}
(q^{d_\alpha h^\p_\alpha})_1\otimes...
\otimes (q^{d_\alpha h^\p_\alpha})_{i-1}\otimes(f^\p_\alpha)_i\otimes
(q^{-d_\alpha h^\p_\alpha})_{i+1}\otimes...
\otimes(q^{-d_\alpha h^\p_\alpha})_{L+1}.
\ea
\ee

{\sf [Theorem 2]}. The chain model defined by the following Hamiltonian
acting in $H_1\otimes ...\otimes H_{L+1}$
is invariant under the quantum algebra $A_q$:
\be\label{33}
H_q=\sum_{i=1}^L(\Delta^\p \Fb(C_q(e^\prime)))_{i,i+1}.
\ee

{\sf [Proof]}. Using (\ref{32}) and (\ref{25}) we get
$$
\ba{rcl}
[H_q,H^\p_\alpha]
&=&\displaystyle\sum_{i=1}^L\left[(\Delta^\p
\Fb(C_q(e^\prime)))_{i,i+1},\displaystyle\sum_{j=1}^{L+1}(h^\p_\alpha)_j\right]\\[5mm]
&=&\displaystyle\sum_{i=1}^L\left[(\Delta^\p
\Fb(C_q(e^\prime)))_{i,i+1},(h^\p_\alpha)_i+(h^\p_\alpha)_{i+1}\right]\\[5mm]
&=&\displaystyle\sum_{i=1}^L\left[(\Delta^\p
\Fb(C_q(e^\prime))),\Delta^\p (h^\p_\alpha)\right]_{i,i+1}=0.
\ea
$$

From formulae (\ref{32}) and (\ref{31}) we have
$$
\ba{rcl}
[H_q,E^\p_\alpha]
&=&\left[\displaystyle\sum_{i=1}^L(\Delta^\p
\Fb(C_q(e^\prime)))_{i,i+1},\right.\\[5mm]
&&\left.\displaystyle\sum_{j=1}^{L+1}
(q^{d_\alpha h^\p_\alpha})_1\otimes...
\otimes (q^{d_\alpha h^\p_\alpha})_{j-1}\otimes(e^\p_\alpha)_j\otimes
(q^{-d_\alpha h^\p_\alpha})_{j+1}\otimes...(q^{-d_\alpha h^\p_\alpha})_{L+1}\right]\\[5mm]
&=&\displaystyle\sum_{i=1}^L\left[(\Delta^\p
\Fb(C_q(e^\prime)))_{i,i+1},
(e^\p_\alpha)_i\otimes (q^{-d_\alpha h^\p_\alpha})_{i+1}+
(q^{d_\alpha h^\p_\alpha})_i\otimes(e^\p_\alpha)_{i+1}\right].
\ea
$$
Using formulae (\ref{26}) and (\ref{29}) we get
$$
[H_q,E^\p_\alpha]=\sum_{i=1}^L\left[(\Delta^\p
\Fb(C_q(e^\prime))),\Delta^\p (e^\p_\alpha)\right]_{i,i+1}=0.
$$

Similarly we have
$$
\ba{rcl}
[H_q,F^\p_\alpha]
&=&\left[\displaystyle\sum_{i=1}^L(\Delta^\p
\Fb(C_q(e^\prime)))_{i,i+1},\right.\\[5mm]
&&\left.\displaystyle\sum_{j=1}^{L+1}
(q^{d_\alpha h^\p_\alpha})_1\otimes...
\otimes (q^{d_\alpha h^\p_\alpha})_{j-1}\otimes(f^\p_\alpha)_j\otimes
(q^{-d_\alpha h^\p_\alpha})_{j+1}\otimes...(q^{-d_\alpha h^\p_\alpha})_{L+1}\right]\\[5mm]
&=&\displaystyle\sum_{i=1}^L\left[(\Delta^\p
\Fb(C_q(e^\prime)))_{i,i+1},
(e^\p_\alpha)_i\otimes (q^{-d_\alpha h^\p_\alpha})_{i+1}+
(q^{d_\alpha h^\p_\alpha})_i\otimes(f^\p_\alpha)_{i+1}\right]\\[5mm]
&=&\displaystyle\sum_{i=1}^L\left[(\Delta^\p
\Fb(C_q(e^\prime))),\Delta^\p (f^\p_\alpha)\right]_{i,i+1}=0.
\ea
$$
Therefore $H_q$ commutes with the generators of $A_q$ for the chain.
\hfill $\rule{3mm}{3mm}$

The Hamiltonian (\ref{33}) stands for a system with 
nearest neighbours interactions. Generally by using
the coproduct operator $\Delta^\p$ we can construct models
with $N+1(N\leq L)$ sites interactions:
\be\label{34}
H^N_q=\sum_{i=1}^{L-N+1} \Fb(\Delta^{\p N} C_q(e^\p))_{i,i+1,...,i+N},
\ee

Taking into account the relation
$$
\Delta^{\p N}\left([\Fb(C_q(e^\prime)\right), q^{\pm d_\alpha h_\alpha^\p}])=
[\Delta^{\p N} \Fb(C_q(e^\prime)), \underbrace{q^{\pm d_\alpha h_\alpha^\p}
\otimes...\otimes q^{\pm d_\alpha h_\alpha^\p}}_{N~times}]=0,
$$
we can prove that the Hamiltonian $H^N_q$ has the symmetry 
of the algebra $A_q$. In fact:
$$
\ba{rcl}
[H_q^N,H^\p_\alpha]
&=&\displaystyle\sum_{i=1}^{L-N+1}\left[(\Delta^{\p N}
\Fb(C_q(e^\prime)))_{i,i+1,...,i+N},
\displaystyle\sum_{j=1}^{L+1}(h^\p_\alpha)_j\right]\\[5mm]
&=&\displaystyle\sum_{i=1}^{L-N+1}\left[(\Delta^{\p N}
\Fb(C_q(e^\prime)))_{i,i+1,...,i+N},
\displaystyle\sum_{j=i}^{i+N}(h^\p_\alpha)_j\right]\\[5mm]
&=&\displaystyle\sum_{i=1}^{L-N+1}\left[(\Delta^{\p N}
\Fb(C_q(e^\prime))),\Delta^{\p N}
(h^\p_\alpha)\right]_{i,i+1,...,i+N}=0;
\ea
$$

$$
\ba{rcl}
[H_q^N,E^\p_\alpha]
&=&\left[\displaystyle\sum_{i=1}^{L-N+1}(\Delta^{\p N}
\Fb(C_q(e^\prime)))_{i,i+1,...,i+N},\right.\\[5mm]
&&\left.\displaystyle\sum_{j=i}^{i+N}
(q^{d_\alpha h^\p_\alpha})_i\otimes...
\otimes (q^{d_\alpha h^\p_\alpha})_{j-1}\otimes(e^\p_\alpha)_j\otimes
(q^{-d_\alpha h^\p_\alpha})_{j+1}\otimes...\otimes(q^{-d_\alpha h^\p_\alpha})_{i+N}
\right]\\[5mm]
&=&\displaystyle\sum_{i=1}^{L-N+1}\left[(\Delta^{\p N}
\Fb(C_q(e^\prime))),\Delta^{\p N}
(e^\p_\alpha)\right]_{i,i+1,...,i+N}=0;
\ea
$$

$$
\ba{rcl}
[H_q^N,F^\p_\alpha]
&=&\left[\displaystyle\sum_{i=1}^{L-N+1}(\Delta^{\p N}
\Fb(C_q(e^\prime)))_{i,i+1,...,i+N},\right.\\[5mm]
&&\left.\displaystyle\sum_{j=i}^{i+N}
(q^{d_\alpha h^\p_\alpha})_i\otimes...
\otimes (q^{d_\alpha h^\p_\alpha})_{j-1}\otimes(f^\p_\alpha)_j\otimes
(q^{-d_\alpha h^\p_\alpha})_{j+1}\otimes...\otimes(q^{-d_\alpha h^\p_\alpha})_{i+N}
\right]\\[5mm]
&=&\displaystyle\sum_{i=1}^{L-N+1}\left[(\Delta^{\p N}
\Fb(C_q(e^\prime))),\Delta^{\p N} (f^\p_\alpha)\right]_{i,i+1,...,i+N}=0.
\ea
$$

The Hamiltonian system (\ref{33}) is expressed by the quantum
algebraic generators $e^\p=(h^\p_\alpha,e^\p_\alpha,f^\p_\alpha)$. 
Assume now that $e\to e^\prime(e)$ is an
algebraic map from $A$ to $A_q$ (we remark that for rank one algebras, 
both classical and quantum
algebraic maps can be discussed in terms of the two dimensional
manifolds related to the algebras, see  \cite{fa}). We then have
\be\label{35}
H^N_q=\sum_{i=1}^{L-N+1} \Fb(\Delta^{\p N}
C_q(e^\p(e)))_{i,i+1,...,i+N}.
\ee
In this way we obtain Hamiltonian systems having quantum algebraic symmetry
but expressed in terms of the usual Lie algebraic generators $\{e_{\alpha}\}$.

\section{Integrable Models with $A_n$ Symmetry}

\subsection{Quantum Yang-Baxter Equation}

The quantum Yang-Baxter equation (QYBE) \cite{yang} is the master 
equation for integrable models in statistical mechanics.
It plays an important role in a variety of problems
in theoretical physics such as exactly soluble models (like the six and
eight vertex models) in statistical mechanics \cite{baxter}, 
integrable model field theories \cite{3}, exact S-matrix theoretical
models \cite{4}, two dimensional field theories involving fields with 
intermediate statistics \cite{5}, conformal field theory
and quantum groups \cite{pressley}. In this section we will investigate the
integrability of the chain models having a certain algebraic 
symmetry constructed in section 2. We also
present a series of solutions of the QYBE from the construction of
integrable models.

Let $V$ be a complex vector space and $R$ be the solution of QYBE without
spectral parameters, see e.g. \cite{pressley}. Then $R$ 
takes values in $End_{\Cb}(V\otimes V)$. The QYBE is
\begin{equation}\label{36}
R_{12}R_{13}R_{23}=R_{23}R_{13}R_{12}.
\end{equation}
Here $R_{ij}$ denotes the matrix on the complex vector space 
$V\otimes V\otimes V$, acting as $R$ on the $i$-th and the 
$j$-th components and as the identity on the other components.

Let $\check{R}=Rp$, $p$ as in (\ref{6}).
Then the QYBE (\ref{36}) becomes
\be\label{37}
\check{R}_{12}\check{R}_{23}\check{R}_{12}=
\check{R}_{23}\check{R}_{12}\check{R}_{23},
\ee
where  $\check{R}_{12}=\check{R}\otimes\1$, $\check{R}_{23}=\1\otimes
\check{R}$ and $\1$ is the identity operator on $V$. 

A chain model with nearest neighbours interactions having a (quantum mechanical)
Hamiltonian of the form
\be\label{38}
H=\sum_{i=1}^L({\cal H})_{i,i+1}
\ee
is said to be integrable if the rank-two tensor operator
${\cal H}$ satisfies the QYBE relation (\ref{37}), i.e.,
\be\label{39}
({\cal H})_{12}({\cal H})_{23}({\cal H})_{12}=
({\cal H})_{23}({\cal H})_{12}({\cal H})_{23},
\ee
where 
$$
({\cal H})_{12}={\cal H}\otimes \1,~~~
({\cal H})_{23}=\1 \otimes {\cal H}.
$$

The Hamiltonian system (\ref{38}) satisfying relation (\ref{39}) can in
principle be exactly solved by the algebraic Bethe Ansatz method, see e.g.
\cite{faddeev}.

\subsection{Integrable $A_n$ Symmetric Chain Models}

The integrability of the models having a certain algebraic symmetry
presented in section 2 depends on the detailed representation of the
corresponding symmetry algebra. In the following we investigate the integrability 
of chain models with nearest neighbours interactions and Lie
algebraic symmetry $A_n$.

Let $(a_{\alpha\beta})$ be the Cartan matrix of the $A_n$ algebra. In
the Chevalley basis the algebra $A_n$ is spanned by the generators 
$\{h_\alpha,e_\alpha,f_\alpha\}$, $\alpha=1,2,...,n$, with the following
algebraic relations:
\be\label{40}
\ba{l}
[h_\alpha,h_\beta]=0,\\[3mm]
[h_\alpha,e_\beta]=a_{\alpha\beta}e_\beta,\\[3mm]
[h_\alpha,f_\beta]=-a_{\alpha\beta}f_\beta,\\[3mm]
[e_\alpha,f_\beta]=\delta_{\alpha\beta}h_\alpha,
\ea
\ee
together with the generators with respect to non simple roots,
\be\label{41}
e_{\alpha...\beta\gamma}=[e_\alpha,...,[e_\beta,e_\gamma]...],~~~
f_{\alpha...\beta\gamma}=[f_\alpha,...,[f_\beta,f_\gamma]...].
\ee

Let $E_{\alpha\beta}$ be an $(n+1)\times (n+1)$ matrix such that 
$(E_{\alpha\beta})_{\gamma\delta}=\delta_{\alpha\gamma}\delta_{\beta\delta}$,
i.e., the only non zero element of the matrix $E_{\alpha\beta}$ is $1$
at row $\alpha$ and column $\beta$. Hence
\be\label{42}
E_{\alpha\beta}E_{\gamma\delta}
=\delta_{\beta\gamma}E_{\alpha\delta}
\ee
and 
$$
[E_{\alpha\beta},E_{\gamma\delta}]=\delta_{\beta\gamma}E_{\alpha\delta}
-\delta_{\delta\alpha}E_{\beta\gamma}.
$$

For the fundamental representation we take the basis of the algebra $A_n$ as
\be\label{43}
\ba{l}
h_\alpha =E_{\alpha\alpha}-E_{\alpha+1,\alpha+1},~~~\alpha=1,2,...,n\\[3mm]
\left.\ba{l}
e=\{E_{\alpha\beta}\}\\[3mm]
f=\{E_{\beta\alpha}\}
\ea
~~\right\}~~~~\beta>\alpha=1,2,...,n
\ea
\ee
Both $\{e_\alpha\}$ and $\{f_\alpha\}$ have a total of $n(n+1)/2$ generators.

With respect to the basis (\ref{43}), the Casimir operator of the 
algebra $A_n$ is given by
\be\label{44}
\ba{rcl}
C_{A_n}&=&(n+1)\displaystyle\sum_{\alpha=1}^{n(n+1)/2}
(e_\alpha f_\alpha+f_\alpha e_\alpha)
+\displaystyle\sum_{\alpha=1}^{n}\alpha(n+1-\alpha)h_\alpha^2\\[5mm]
&&+\displaystyle\sum_{\alpha=1}^{n}\displaystyle\sum_{\beta=1}^{n-\alpha}
2\alpha(n+1-\alpha-\beta)h_\alpha h_{\alpha+\beta}-a,
\ea
\ee
where $a$ is an arbitrary real constant.

The coproduct operator $\Delta$ is given by
\be\label{45}
\ba{ll}
\,~\Delta(\1)=\1\otimes\1\\[3mm]
\,~\Delta(h_\alpha)=h_\alpha\otimes\1+\1\otimes h_\alpha,
&~~~\alpha=1,2,...,n\\[3mm]
\left.\ba{l}
\Delta(e_\beta)=e_\beta\otimes\1+\1\otimes e_\beta\\[3mm]
\Delta(f_\beta)=f_\beta\otimes\1+\1\otimes f_\beta
\ea
~~\right\}&~~~~\beta=1,2,...,n(n+1)/2,
\ea
\ee
where the identity operator $\1$ is the $(n+1)\times (n+1)$ ~identity
matrix.

By (\ref{44}) and (\ref{45}) we have
\be\label{46}
\ba{ll}
\Delta C_{A_n}=&C_{A_n}\otimes\1 +\1\otimes C_{A_n}-a\1\otimes\1 \\[4mm]
&+(n+1)\displaystyle\sum_{\alpha=1}^{n(n+1)/2}
(e_\alpha\otimes f_\alpha+f_\alpha\otimes e_\alpha)
+\displaystyle\sum_{\alpha=1}^{n}\alpha(n+1-\alpha)h_\alpha\otimes h_\alpha\\[5mm]
&+\displaystyle\sum_{\alpha=1}^{n}\displaystyle\sum_{\beta=1}^{n-\alpha}
\alpha(n+1-\alpha-\beta)(h_\alpha\otimes h_{\alpha+\beta}
+h_{\alpha+\beta}\otimes h_{\alpha}).
\ea
\ee

It is easy to check that under the representation (\ref{43}) $C_{A_n}$
is equal to $n(n+2)\1$. Therefore the
sum of the first two terms on the right hand side of (\ref{46}) is
$2n(n+2)\1\times\1$. In the following we take $a$ in (\ref{46}) to be
$2n(n+2)$ so that the terms that are
proportional to $(n+1)^2\times (n+1)^2$ identity matrix 
will disappear in (\ref{46}).

From (\ref{43}) and (\ref{46}) we have
\be\label{47}
\ba{ll}
\Delta C_{A_n}=&
(n+1)\displaystyle\sum_{\alpha\neq\beta=1}^{n+1}E_{\alpha\beta}\otimes
E_{\beta\alpha}\\[5mm]
&+\displaystyle\sum_{\alpha=1}^{n}\alpha(n+1-\alpha)
(E_{\alpha\alpha}-E_{\alpha+1,\alpha+1})\otimes 
(E_{\alpha\alpha}-E_{\alpha+1,\alpha+1})\\[5mm]
&+\displaystyle\sum_{\alpha=1}^{n}\displaystyle\sum_{\beta=1}^{n-\alpha}
\alpha(n+1-\alpha-\beta)[(E_{\alpha\alpha}-E_{\alpha+1,\alpha+1})
\otimes (E_{\alpha+\beta,\alpha+\beta}-
E_{\alpha+\beta+1,\alpha+\beta+1})\\[5mm]
&+ (E_{\alpha+\beta,\alpha+\beta}-E_{\alpha+\beta+1,\alpha+\beta+1})
\otimes (E_{\alpha\alpha}-E_{\alpha+1,\alpha+1})].
\ea
\ee

$\Delta C_{A_n}$ in (\ref{47}) is an $(n+1)^2\times (n+1)^2$ matrix. Its
matrix representation is
\be\label{48}
\ba{ll}
(\Delta C_{A_n})_{\alpha\beta}=
&\delta_{\alpha\beta}[(n+1)\delta_{\alpha,l(n+1)+l+1}-1]\\[3mm]
&+(n+1)[\delta_{\alpha,j(n+2)+k+2}\delta_{\beta,(j+1)(n+2)+k(n+1)}\\[3mm]
&+\delta_{\beta,j(n+2)+k+2}\delta_{\alpha,(j+1)(n+2)+k(n+1)}],
\ea
\ee
where $\alpha,\beta=1,2,...,(n+1)^2$, $l=0,1,...,n$, $j=0,1,...,n-1$, 
$k=0,1,...,n-j-1$, $\delta_{\alpha,j(n+2)+k+2}=0$ if 
$\alpha\neq j(n+2)+k+2$ for all possible values of $j$ and $k$. 
Here we give explicitly, as examples, the matrix representations of $\Delta C_{A_n}$
for $n=1,2,3$:

\be\label{49}
\Delta C_{A_1}=\left(
\ba{cccc}
1&\cdot&\cdot&\cdot\\[2mm]
\cdot&-1&2&\cdot\\[2mm]
\cdot&2&-1&\cdot\\[2mm]
\cdot&\cdot&\cdot&1
\ea
\right),
\ee

\be\label{50}
\Delta C_{A_2}=\left(
\ba{ccccccccc}
2&\cdot&\cdot&\cdot&\cdot&\cdot&\cdot&\cdot&\cdot\\[2mm]
\cdot&-1&\cdot&3&\cdot&\cdot&\cdot&\cdot&\cdot\\[2mm]
\cdot&\cdot&-1&\cdot&\cdot&\cdot&3&\cdot&\cdot\\[2mm]
\cdot&3&\cdot&-1&\cdot&\cdot&\cdot&\cdot&\cdot\\[2mm]
\cdot&\cdot&\cdot&\cdot&2&\cdot&\cdot&\cdot&\cdot\\[2mm]
\cdot&\cdot&\cdot&\cdot&\cdot&-1&\cdot&3&\cdot\\[2mm]
\cdot&\cdot&3&\cdot&\cdot&\cdot&-1&\cdot&\cdot\\[2mm]
\cdot&\cdot&\cdot&\cdot&\cdot&3&\cdot&-1&\cdot\\[2mm]
\cdot&\cdot&\cdot&\cdot&\cdot&\cdot&\cdot&\cdot&2
\ea
\right),
\ee

\be\label{51}
\Delta C_{A_3}=\left(
\ba{cccccccccccccccc}
3&\cdot&\cdot&\cdot&\cdot&\cdot&\cdot&\cdot&\cdot&\cdot&\cdot&\cdot&
\cdot&\cdot&\cdot&\cdot\\[2mm]
\cdot&-1&\cdot&\cdot&4&\cdot&\cdot&\cdot&\cdot&\cdot&\cdot&\cdot&
\cdot&\cdot&\cdot&\cdot\\[2mm]
\cdot&\cdot&-1&\cdot&\cdot&\cdot&\cdot&\cdot&4&\cdot&\cdot&\cdot&\cdot&
\cdot&\cdot&\cdot\\[2mm]
\cdot&\cdot&\cdot&-1&\cdot&\cdot&\cdot&\cdot&\cdot&\cdot&\cdot&
\cdot&4&\cdot&\cdot&\cdot\\[2mm]
\cdot&4&\cdot&\cdot&-1&\cdot&\cdot&\cdot&\cdot&\cdot&\cdot&\cdot&
\cdot&\cdot&\cdot&\cdot\\[2mm]
\cdot&\cdot&\cdot&\cdot&\cdot&3&\cdot&\cdot&\cdot&\cdot&\cdot&\cdot&
\cdot&\cdot&\cdot&\cdot\\[2mm]
\cdot&\cdot&\cdot&\cdot&\cdot&\cdot&-1&\cdot&\cdot&4&\cdot&\cdot&\cdot&
\cdot&\cdot&\cdot\\[2mm]
\cdot&\cdot&\cdot&\cdot&\cdot&\cdot&\cdot&-1&\cdot&\cdot&\cdot&\cdot&
\cdot&4&\cdot&\cdot\\[2mm]
\cdot&\cdot&4&\cdot&\cdot&\cdot&\cdot&\cdot&-1&\cdot&\cdot&\cdot&
\cdot&\cdot&\cdot&\cdot\\[2mm]
\cdot&\cdot&\cdot&\cdot&\cdot&\cdot&4&\cdot&\cdot&-1&\cdot&\cdot&
\cdot&\cdot&\cdot&\cdot\\[2mm]
\cdot&\cdot&\cdot&\cdot&\cdot&\cdot&\cdot&\cdot&\cdot&\cdot&3&\cdot&
\cdot&\cdot&\cdot&\cdot\\[2mm]
\cdot&&\cdot&\cdot&\cdot&\cdot&\cdot&\cdot&\cdot&\cdot&\cdot&-1&
\cdot&\cdot&4&\cdot\\[2mm]
\cdot&\cdot&\cdot&4&\cdot&\cdot&\cdot&\cdot&\cdot&\cdot&\cdot&
\cdot&-1&\cdot&\cdot&\cdot\\[2mm]
\cdot&\cdot&\cdot&\cdot&\cdot&\cdot&\cdot&4&\cdot&\cdot&\cdot&
\cdot&\cdot&-1&\cdot&\cdot\\[2mm]
\cdot&\cdot&\cdot&\cdot&\cdot&\cdot&\cdot&\cdot&\cdot&\cdot&
\cdot&4&\cdot&\cdot&-1&\cdot\\[2mm]
\cdot&\cdot&\cdot&\cdot&\cdot&\cdot&\cdot&\cdot&\cdot&\cdot&
\cdot&\cdot&\cdot&\cdot&\cdot&3
\ea
\right),
\ee
where for simplicity $\cdot$ stands for $0$.

{\sf [Lemma 1].} $\Delta C_{A_n}$ satisfies the following relation
\be\label{53}
(\Delta C_{A_n})^2+2\Delta C_{A_n}-n(n+2)\1\otimes\1=0.
\ee

{\sf [Proof].} From (\ref{48}) we have
$$
\ba{rcl}
[(\Delta C_{A_n})^2]_{\alpha\gamma}
&=&\displaystyle\sum_{\beta=1}^{(n+1)^2}
(\Delta C_{A_n})_{\alpha\beta}(\Delta C_{A_n})_{\beta\gamma}\\[3mm]
&=&\displaystyle\sum_{\beta=1}^{(n+1)^2}
[\delta_{\alpha\beta}((n+1)\delta_{\alpha,l(n+1)+l+1}-1)
+(n+1)(\delta_{\alpha,j(n+2)+k+2}\delta_{\beta,(j+1)(n+2)+k(n+1)}\\[3mm]
&&+\delta_{\beta,j(n+2)+k+2}\delta_{\alpha,(j+1)(n+2)+k(n+1)})]\cdot\\[3mm]
&&[\delta_{\beta\gamma}((n+1)\delta_{\beta,l^\p (n+1)+l^\p +1}-1)
+(n+1)(\delta_{\beta,j^\p (n+2)+k^\p +2}
\delta_{\gamma,(j^\p +1)(n+2)+k^\p (n+1)}\\[3mm]
&&+\delta_{\gamma,j^\p (n+2)+k^\p +2}
\delta_{\beta,(j^\p +1)(n+2)+k^\p (n+1)})]\\[3mm]
&=&\delta_{\alpha\gamma}[(n+1)^2\delta_{\alpha,l(n+1)+l+1}
\delta_{\gamma,l^\p (n+1)+l^\p +1}]\\[3mm]
&&-(n+1)(\delta_{\alpha,l(n+1)+l+1}+\delta_{\gamma,l^\p (n+1)+l^\p +1})+1]
\\[3mm]
&&-2(n+1)[\delta_{\alpha,j(n+2)+k+2}\delta_{\gamma,(j+1)(n+2)+k(n+1)}
+\delta_{\gamma,j(n+2)+k+2}\delta_{\alpha,(j+1)(n+2)+k(n+1)}]\\[3mm]
&&+(n+1)^2[\delta_{\alpha,j(n+2)+k+2}\delta_{\gamma,j(n+2)+k+2}
+\delta_{\alpha,(j+1)(n+2)+k(n+1)}\delta_{\gamma,(j+1)(n+2)+k(n+1)}]\\[3mm]
&=&\delta_{\alpha\gamma}[(n-1)^2\delta_{\alpha,l(n+1)+l+1}]\\[3mm]
&&-2(n+1)[\delta_{\alpha,j(n+2)+k+2}\delta_{\gamma,(j+1)(n+2)+k(n+1)}
+\delta_{\gamma,j(n+2)+k+2}\delta_{\alpha,(j+1)(n+2)+k(n+1)}]\\[3mm]
&&+(n+1)^2\delta_{\alpha\gamma}
[\delta_{\alpha,j(n+2)+k+2}+\delta_{\alpha,(j+1)(n+2)+k(n+1)}]\\[3mm]
&=&-2(\Delta C_{A_n})_{\alpha\gamma}+
(n+1)^2\delta_{\alpha\gamma}
[\delta_{\alpha,j(n+2)+k+2}+\delta_{\alpha,(j+1)(n+2)+k(n+1)}]\\[3mm]
&&+(n+1)^2\delta_{\alpha\gamma}\delta_{\alpha,l(n+1)+l+1}-\delta_{\alpha\gamma}
\\[3mm]
&=&-2(\Delta C_{A_n})_{\alpha\gamma}+(n+1)^2\delta_{\alpha\gamma}
-\delta_{\alpha\gamma}\\[3mm]
&=&-2(\Delta C_{A_n})_{\alpha\gamma}+n(n+2)\delta_{\alpha\gamma},
\ea
$$
where the identity
\be\label{identity}
\delta_{\alpha,l(n+1)+l+1}+
\delta_{\alpha,j(n+2)+k+2}+\delta_{\alpha,(j+1)(n+2)+k(n+1)}=1,
\ee
$l=0,1,...,n$, $j=0,1,...,n-1$, $k=0,1,...,n-j-1$, has been used.
\hfill $\rule{3mm}{3mm}$

{\sf [Lemma 2]}. The coproduct of the $A_n$ Casimir operator $\Delta
C_{A_n}$ has the following properties:
\be\label{l21}
\ba{l}
(\Delta C_{A_n}\otimes\1)(\1\otimes\Delta C_{A_n})(\Delta C_{A_n}\otimes\1)
\\[3mm]
-n[(\1\otimes\Delta C_{A_n})(\Delta C_{A_n}\otimes\1)+
(\Delta C_{A_n}\otimes\1)(\1\otimes\Delta C_{A_n})]\\[3mm]
+(n^2-1)(\Delta C_{A_n}\otimes\1)+n^2(\1\otimes\Delta C_{A_n})
+n(1-n^2)\1\otimes\1\otimes\1=0
\ea
\ee
and
\be\label{l22}
\ba{l}
(\1\otimes\Delta C_{A_n})(\Delta C_{A_n}\otimes\1)(\1\otimes\Delta C_{A_n})
\\[3mm]
-n[(\1\otimes\Delta C_{A_n})(\Delta C_{A_n}\otimes\1)+
(\Delta C_{A_n}\otimes\1)(\1\otimes\Delta C_{A_n})]\\[3mm]
+(n^2-1)(\1\otimes\Delta C_{A_n})+n^2(\Delta C_{A_n}\otimes\1)
+n(1-n^2)\1\otimes\1\otimes\1=0.
\ea
\ee

{\sf [Proof]}. By using the representation of $\Delta C_{A_n}$ in (\ref{48})
we have
\be\label{dc1}
\ba{rcl}
(\Delta C_{A_n}\otimes\1)_{\alpha\beta}
&=&(\Delta C_{A_n})_{(\alpha-\gamma)/(n+1)+1,(\beta-\gamma)/(n+1)+1}\\[3mm]
&=&\delta_{\alpha\beta}[(n+1)\delta_{\alpha-\gamma,l(n+1)(n+2)}-1]\\[3mm]
&&+(n+1)[\delta_{\alpha-\gamma,(n+1)(j(n+2)+k+1))}\delta_{\beta-\gamma,
(n+1)(j(n+2)+(k+1)(n+1))}\\[3mm]
&&+\delta_{\beta-\gamma,(n+1)(j(n+2)+k+1)}\delta_{\alpha-
\gamma,(n+1)(j(n+2)+(k+1)(n+1))}]
\ea
\ee
and
\be\label{dc2}
\ba{rcl}
(\1\otimes\Delta C_{A_n})_{\alpha\beta}
&=&(\Delta C_{A_n})_{\alpha-(n+1)^2(\gamma^\p-1),\beta-(n+1)^2(\gamma^\p-1)}\\[3mm]
&=&\delta_{\alpha\beta}[(n+1)\delta_{\alpha-(n+1)^2(\gamma^\p-1),l(n+1)+l+1)}-1]\\[3mm]
&&+(n+1)[\delta_{\alpha-(n+1)^2(\gamma^\p-1),j(n+2)+k+2))}\delta_{\beta-(n+1)^2(\gamma^\p-1),
(j+1)(n+2)+k(n+1)}\\[3mm]
&&+\delta_{\beta-(n+1)^2(\gamma^\p-1),j(n+2)+k+2}
\delta_{\alpha-(n+1)^2(\gamma^\p-1),(j+1)(n+2)+k(n+1)}],
\ea
\ee
where $\alpha,\beta=1,...,(n+1)^3$, $l=0,1,...,n$, $j=0,1,...,n-1$ and
$k=0,1,...,n-j-1$ as in formula (\ref{48}), $\gamma=1,...,n+1$ such that 
$(\alpha-\gamma)/(n+1)$ and $(\beta-\gamma)/(n+1)$ in (\ref{dc1}) are integers
and $\gamma^\p=1,...,n+1$ in (\ref{dc2}).

Using the formulae (\ref{dc1}) and (\ref{dc2}) one can get (\ref{l21}) and
(\ref{l22}) from straightforward calculations. \hfill $\rule{3mm}{3mm}$

From Theorem 1 we know that the following Hamiltonian is invariant 
under $A_{n}$
\be\label{54}
H=\sum_{i=1}^L\Fb(\Delta C_{A_n})_{i,i+1}.
\ee
For the given representation (\ref{43}) of $A_n$ the integrability of
(\ref{48}) depends on the form of the entire function $\Fb$. Due to the
relation (\ref{53}) in {\sf Lemma 1}, $(\Delta C_{A_n})^l$, 
$l\geq 2$, can be expressed as $c \Delta C_{A_n}+c^\p\1\otimes\1$ for
some real constants $c$ and $c^\p$. Therefore $\Fb(\Delta C_{A_n})$ 
is a polynomial in $\Delta C_{A_n}$ up to powers of order two.

{\sf [Theorem 3]}. The following $A_n$ invariant Hamiltonian
is integrable

\be\label{55}
\ba{rcl}
H_{A_n}&=&\displaystyle\sum_{i=1}^L({\cal H})_{i,i+1}
=\displaystyle\sum_{i=1}^L(\Delta C_{A_n}+1)_{i,i+1}\\[4mm]
&=&\displaystyle
\sum_{i=1}^L\left[(n+1)\displaystyle\sum_{\alpha=1}^{n(n+1)/2}
((e_\alpha)_i(f_\alpha)_{i+1}+(f_\alpha)_i(e_\alpha)_{i+1})
+\displaystyle\sum_{\alpha=1}^{n}\alpha(n+1-\alpha)
(h_\alpha)_i(h_\alpha)_{i+1}\right.\\[5mm]
&&\left.+\displaystyle\sum_{\alpha=1}^{n}\displaystyle\sum_{\beta=1}^{n-\alpha}
\alpha(n+1-\alpha-\beta)((h_\alpha)_i(h_{\alpha+\beta})_{i+1}
+(h_{\alpha+\beta})_i(h_{\alpha})_{i+1})\right]+L,
\ea
\ee
where ${\cal H}=\Delta C_{A_n}+1$ and the number $1$ should be 
understood as the identity operator, $\1\otimes \1$, on the tensor space
$H_1\otimes...\otimes H_{L+1}$ ($L+1$ is the number of lattice sites of 
the chain). 

{\sf [Proof]}. What we have to prove is that ${\cal H}$ ~satisfies the
QYBE (\ref{39}), i.e.,
$$
({\cal H})_{12}({\cal H})_{23}({\cal H})_{12}=
({\cal H})_{23}({\cal H})_{12}({\cal H})_{23},
$$
where 
\be\label{56}
({\cal H})_{12}=(\Delta C_{A_n}+1)\otimes \1,~~~
({\cal H})_{23}=\1 \otimes(\Delta C_{A_n}+1).
\ee

From (\ref{56}) we have
$$
\ba{l}
({\cal H})_{12}({\cal H})_{23}({\cal H})_{12}\\[4mm]
~=(\1\otimes\Delta C_{A_n}+\Delta C_{A_n}\otimes\1
+(\Delta C_{A_n}\otimes\1)(\1\otimes\Delta C_{A_n})
+\1\otimes\1\otimes\1)({\cal H})_{12}\\[4mm]
\ba{ll}
=&(\1\otimes\Delta C_{A_n})(\Delta C_{A_n}\otimes\1)+
(\Delta C_{A_n}\otimes\1)(\Delta C_{A_n}\otimes\1)\\[4mm]
&+(\Delta C_{A_n}\otimes\1)(\1\otimes\Delta C_{A_n})
(\Delta C_{A_n}\otimes\1)
+(\Delta C_{A_n}\otimes\1)(\1\otimes\Delta C_{A_n})\\[4mm]
&+2\Delta C_{A_n}\otimes\1+\1\otimes\Delta C_{A_n}
+\1\otimes\1\otimes\1
\ea
\ea
$$
and
$$
\ba{l}
({\cal H})_{23}({\cal H})_{12}({\cal H})_{23}\\[4mm]
~=({\cal H})_{23}(\1\otimes\Delta C_{A_n}+\Delta C_{A_n}\otimes\1
+(\Delta C_{A_n}\otimes\1)(\1\otimes\Delta C_{A_n})
+\1\otimes\1\otimes\1)\\[4mm]
\ba{ll}
=&(\1\otimes\Delta C_{A_n})(\1\otimes\Delta C_{A_n})
+(\1\otimes\Delta C_{A_n})(\Delta C_{A_n}\otimes\1)\\[4mm]
&+(\1\otimes\Delta C_{A_n})(\Delta C_{A_n}\otimes\1)
(\1\otimes\Delta C_{A_n})
+(\Delta C_{A_n}\otimes\1)(\1\otimes\Delta C_{A_n})\\[4mm]
&+2(\1\otimes\Delta C_{A_n})+(\Delta C_{A_n}\otimes\1)
+\1\otimes\1\otimes\1.
\ea
\ea
$$
Hence
\be\label{57}
({\cal H})_{12}({\cal H})_{23}({\cal H})_{12}-
({\cal H})_{23}({\cal H})_{12}({\cal H})_{23}=
I+II+III,
\ee
where
$$
\ba{rcl}
I&=&(\Delta C_{A_n}\otimes\1)(\Delta C_{A_n}\otimes\1)
-(\1\otimes\Delta C_{A_n})(\1\otimes\Delta C_{A_n}),\\[4mm]
II&=&(\Delta C_{A_n}\otimes\1)(\1\otimes\Delta C_{A_n})(\Delta C_{A_n}\otimes\1)
-(\1\otimes\Delta C_{A_n})(\Delta C_{A_n}\otimes\1)(\1\otimes\Delta
C_{A_n}),\\[4mm]
III&=&\Delta C_{A_n}\otimes\1-\1\otimes\Delta C_{A_n}.
\ea
$$

Using (\ref{53}) we have
$$
\ba{rcl}
I&=&(\Delta C_{A_n}\otimes\1)(\Delta C_{A_n}\otimes\1)
-(\1\otimes\Delta C_{A_n})(\1\otimes\Delta C_{A_n})\\[4mm]
&=&(\Delta C_{A_n})^2\otimes\1-\1\otimes (\Delta C_{A_n})^2\\[4mm]
&=&-(2\Delta C_{A_n}-n(n+2)\1\otimes\1)\otimes\1
+\1\otimes (2\Delta C_{A_n}-n(n+2)\1\otimes\1)\\[4mm]
&=&-2(\Delta C_{A_n}\otimes\1-\1\otimes\Delta C_{A_n}).
\ea
$$
Therefore
\be\label{58}
I+II=\1\otimes\Delta C_{A_n}-\Delta C_{A_n}\otimes\1.
\ee

By Lemma 2 we get
$$
\ba{rcl}
III&=&(\Delta C_{A_n}\otimes\1)(\1\otimes\Delta C_{A_n})(\Delta C_{A_n}\otimes\1)
-(\1\otimes\Delta C_{A_n})(\Delta C_{A_n}\otimes\1)(\1\otimes\Delta
C_{A_n})\\[4mm]
&=&\Delta C_{A_n}\otimes\1-\1\otimes\Delta C_{A_n}.
\ea
$$
Therefore
$$
({\cal H})_{12}({\cal H})_{23}({\cal H})_{12}-
({\cal H})_{23}({\cal H})_{12}({\cal H})_{23}=
I+II+III=0.
$$
\hfill $\rule{3mm}{3mm}$

\subsection{Temperley-Lieb Algebraic Structures and Representations}

An $L$-state TL algebra is described by the elements
$e_{i}$, $i=1,2,...,L$, satisfying the TL algebraic relations \cite{tl},
\begin{equation}\label{tl}
\begin{array}{l}
e_ie_{i\pm 1}e_i=e_i\,,\\[3mm]
e_ie_j=e_je_i\,,~~~{\sf if~} ~\vert i-j\vert \ge 2\,,
\end{array}
\end{equation}
and
\begin{equation}\label{ei2}
e_i^2=\beta e_i\,,
\end{equation}
where $\beta$ is a complex constant and $i=1,2,\cdots ,L$.

In this section we indicate that there is a TL algebraic 
structure related to the integrable chain model (\ref{55}), 
in the sense that the model gives a representation of the TL algebra.
We suppose that the representation of an $L$-state TL algebra on an $L+1$
chain is of the following form,
\begin{equation}\label{ei}
e_i={\bf 1}_{1}\otimes{\bf 1}_{2}\otimes\cdots\otimes{\bf 1}_{i-1}\otimes
E\otimes{\bf 1}_{i+2}\otimes
\cdots\otimes{\bf 1}_{L+1}\,,
\end{equation}
where $\1$ is the $(n+1)\times (n+1)$ identity matrix as in section 
3.2 and $E$ is a $(n+1)^{2}\times (n+1)^{2}$ matrix. According to
formulae (\ref{ei2}) and (\ref{tl}) $E$ should satisfy
\begin{equation}\label{ee}
E^{2}=\beta E\,.
\end{equation}
\be\label{ey}
\ba{l}
(E\otimes{\bf 1})({\bf 1}\otimes E)(E\otimes{\bf 1})=E\otimes{\bf
1},\\[3mm]
({\bf 1}\otimes E)(E\otimes{\bf 1})({\bf 1}\otimes E)={\bf 1}\otimes E.
\ea
\ee

{\sf [Theorem 4]}. For a given representation of the TL algebra of the form
(\ref{ei}) with $E$ satisfying (\ref{ee}) and (\ref{ey}) we have that
\begin{equation}\label{r}
{\check{R}}=E+\frac{-\beta\pm\sqrt{\beta^{2}-4}}{2}\,
{\bf 1}\otimes {\bf 1}
\end{equation}
is a solution of the QYBE (\ref{37}).

{\sf [Proof]}. For simplicity we set $c=(-\beta\pm\sqrt{\beta^{2}-4})/2$.
Substituting (\ref{r}) into equation 
(\ref{37}) and using relations (\ref{ee}) and (\ref{ey}) we get
$$
\ba{l}
\check{R}_{12}\check{R}_{23}\check{R}_{12}
-\check{R}_{23}\check{R}_{12}\check{R}_{23}\\[3mm]
=(E\otimes{\bf 1})({\bf 1}\otimes E)(E\otimes{\bf 1})
-({\bf 1}\otimes E)(E\otimes{\bf 1})({\bf 1}\otimes E)\\[3mm]
~~~+c(E^2\otimes{\bf 1}-{\bf 1}\otimes E^2)+c^{2}(E\otimes {\bf 1}-{\bf
1}\otimes E)\\[3mm]
=(E\otimes{\bf 1})({\bf 1}\otimes E)(E\otimes{\bf 1})
-({\bf 1}\otimes E)(E\otimes{\bf 1})({\bf 1}\otimes E)
+(c\beta+c^2)(E\otimes {\bf 1}-{\bf
1}\otimes E)\\[3mm]
=(c\beta+c^2+1)(E\otimes {\bf 1}-{\bf
1}\otimes E)=0.
\ea
$$
\hfill $\rule{3mm}{3mm}$

In general however the converse does not hold, i.e.,
for a given solution ${\check{R}}$ of the QYBE
(\ref{37}), there does not necessarily exit a TL algebraic 
representation of the form 
(\ref{ei}) with $E=a{\check{R}}+b$ satisfying (\ref{ee}) and (\ref{ey}) for
any constants $a$ and $b$. Nevertheless the solutions ${\cal H}$ of the
QYBE in our $A_n$ symmetric integrable model (\ref{55}) do give rise to 
TL algebraic representations in the following sense:

{\sf [Theorem 5]}. The following $(n+1)^2\times (n+1)^2$ matrix
\be\label{ae}
E=-\frac{{\cal H}}{n+1}+\1\otimes\1
\ee
gives the $L$-state TL algebraic representation (\ref{ei}) with
$\beta=2$.

{\sf [Proof]}. What we should check is that $E$ in (\ref{ae})
satisfies equations (\ref{ee}) and (\ref{ey}). By Lemma 1 we have
$$
\ba{rcl}
E^2&=&(-\displaystyle\frac{{\cal H}}{n+1}+\1\otimes\1)^2
=(-\displaystyle\frac{\Delta C_{A_n}+1\otimes\1}{n+1}+\1\otimes\1)^2\\[3mm]
&=&\displaystyle\frac{(\Delta C_{A_n})^2-2n \Delta C_{A_n}+n^2\1\otimes\1}{(n+1)^2}\\[3mm]
&=&\displaystyle\frac{(-2(n+1)\Delta C_{A_n}+2n(n+1)\1\otimes\1}{(n+1)^2}\\[3mm]
&=&\beta E=2E,
\ea
$$
i.e., $\beta=2$.

From Lemma 1 and (\ref{l21}) in Lemma 2 we get
$$
\ba{l}
(E\otimes{\bf 1})({\bf 1}\otimes E)(E\otimes{\bf 1})\\[4mm]
=(\1\otimes\1\otimes\1-\displaystyle\frac{{\cal H}\otimes\1}{n+1})
(\1\otimes\1\otimes\1-\displaystyle\frac{\1\otimes{\cal H}}{n+1})
(\1\otimes\1\otimes\1-\displaystyle\frac{{\cal H}\otimes\1}{n+1})\\[4mm]
=\displaystyle\frac{-1}{(n+1)^3}[
(\Delta C_{A_n}\otimes\1)(\1\otimes\Delta C_{A_n})(\Delta C_{A_n}\otimes\1)\\[4mm]
~~~-n((\Delta C_{A_n}\otimes\1)(\1\otimes\Delta C_{A_n})+
(\1\otimes\Delta C_{A_n})(\Delta C_{A_n}\otimes\1))\\[4mm]
~~~-n(\Delta C_{A_n})^2\otimes\1+2n^2\Delta C_{A_n}\otimes\1
+n^2\1\otimes\Delta C_{A_n}-n^3\1\otimes\1\otimes\1]\\[4mm]
=\displaystyle\frac{-1}{(n+1)^3}[
(\Delta C_{A_n}\otimes\1)(\1\otimes\Delta C_{A_n})(\Delta C_{A_n}\otimes\1)\\[4mm]
~~~-n((\Delta C_{A_n}\otimes\1)(\1\otimes\Delta C_{A_n})+
(\1\otimes\Delta C_{A_n})(\Delta C_{A_n}\otimes\1))\\[4mm]
~~~+2n(n+1)\Delta C_{A_n}\otimes\1
+n^2\1\otimes\Delta C_{A_n}-2(n^3+n^2)\1\otimes\1\otimes\1]\\[4mm]
=\displaystyle\frac{-1}{(n+1)^3}[(n+1)^2\Delta C_{A_n}\otimes\1
-n(n+1)^2\1\otimes\1\otimes\1]\\[4mm]
=\displaystyle\frac{n\1\otimes\1\otimes\1-\Delta C_{A_n}\otimes\1}{(n+1)}
=E\otimes{\bf 1}.
\ea
$$

By using Lemma 1 and formula (\ref{l22}) in Lemma 2 we then conclude
that
$$
\ba{l}
({\bf 1}\otimes E)(E\otimes{\bf 1})({\bf 1}\otimes E)\\[4mm]
=(\1\otimes\1\otimes\1-\displaystyle\frac{\1\otimes{\cal H}}{n+1})
(\1\otimes\1\otimes\1-\displaystyle\frac{{\cal H}\otimes\1}{n+1})
(\1\otimes\1\otimes\1-\displaystyle\frac{\1\otimes{\cal H}}{n+1})\\[4mm]
=\displaystyle\frac{-1}{(n+1)^3}[(\1\otimes\Delta C_{A_n})
(\Delta C_{A_n}\otimes\1)(\1\otimes\Delta C_{A_n})\\[4mm]
~~~-n((\Delta C_{A_n}\otimes\1)(\1\otimes\Delta C_{A_n})+
(\1\otimes\Delta C_{A_n})(\Delta C_{A_n}\otimes\1))\\[4mm]
~~~-n\1\otimes(\Delta C_{A_n})^2+2n^2\1\otimes\Delta C_{A_n}
+n^2\Delta C_{A_n}\otimes\1-n^3\1\otimes\1\otimes\1]\\[4mm]
=\displaystyle\frac{-1}{(n+1)^3}[
(\Delta C_{A_n}\otimes\1)(\1\otimes\Delta C_{A_n})(\Delta C_{A_n}\otimes\1)\\[4mm]
~~~-n((\Delta C_{A_n}\otimes\1)(\1\otimes\Delta C_{A_n})+
(\1\otimes\Delta C_{A_n})(\Delta C_{A_n}\otimes\1))\\[4mm]
~~~+2n(n+1)\1\otimes\Delta C_{A_n}
+n^2\Delta C_{A_n}\otimes\1-2(n^3+n^2)\1\otimes\1\otimes\1]\\[4mm]
=\displaystyle\frac{-1}{(n+1)^3}[(n+1)^2\1\otimes\Delta C_{A_n}
-n(n+1)^2\1\otimes\1\otimes\1]\\[4mm]
=\displaystyle\frac{n\1\otimes\1\otimes\1-\1\otimes\Delta C_{A_n}}{(n+1)}
=\1\otimes E.
\ea
$$
\hfill $\rule{3mm}{3mm}$

From (\ref{ae}) we see that the Hamiltonian of the $A_n$ symmetric
integrable chain model (\ref{55}) can be expressed by the TL algebraic
elements
\be
H_{A_n}=\displaystyle\sum_{i=1}^L({\cal H})_{i,i+1}
=\displaystyle\sum_{i=1}^L(n+1)(-E+1)_{i,i+1}
=\displaystyle\sum_{i=1}^L(n+1)e_i+(n+1)L,
\ee
with $e_i$ as in (\ref{ei}) and $E$ as in (\ref{ae}). Hence instead of the algebraic
Bethe Ansatz method, the energy spectrum of $H_{A_n}$ can also be
studied by using the properties of the TL algebra \cite{levy} (for the
case of Heisenberg spin chain model, $n=1$, see \cite{mahou}).

\section{Integrable Models and Stationary Markov Chains}

\subsection{Stationary Markov Chains}

We first briefly recall some concepts of the theory of Markov chains (for a detailed
mathematical description of Markov chains, we refer to \cite{markov}).
Let $\Omega$ denote the sample space (the set of all possible outcomes
of an experiment). If $\Omega$ is a
finite or countably infinite sample space and if $P$ is a probability
measure defined on the $\sigma$-algebra of all subsets of
$\Omega$, then the pair $(\Omega,P)$ is called a probability
space. A subset $A$ of $\Omega$ is then called an event with 
probability $P(A)$.

A function $X\equiv X(\omega)$, $\omega\in \Omega$, 
that maps a sample space into the real numbers 
is called a random variable. A stochastic process is a family
$(X_t)_{t\in I}$, $I$ a certain index set, of random
variables defined on some sample space $\Omega$. If $I$ is countable, i.e.,
$I\in \Nb$, the process is denoted by $X_1,X_2,...$ and
called a discrete-time process. If $I= \Rb_+$, then 
the process is denoted by $\{X_t\}_{t\ge 0}$ and called a
continuous-time process.

The ranges of $X$ (a subset of real numbers) is called the state space.
In what follows we consider the case where the state space $S$
is countable or finite. In this case the related stochastic process is
called a (stochastic or random) chain.

Let $(\Omega,P)$ be a probability space in above sense
and $E,F$ be two subsets of $\Omega$. We denote by $P(E\vert F)$
the (conditional) probability of $E$ given that
$F$ has occurred. A discrete-time stochastic process $\{X_i\}$,
$i=1,2,...$ with state space $S=\Nb$ is said to satisfy the
Markov property if for every $l$ and all states $i_1,i_2,...,i_l$ it is
true that 
$$
P[X_l=i_l\vert X_{l-1}=i_{l-1},X_{l-2}=i_{l-2},...,X_{1}=i_{1}]=
P[X_l=i_l\vert X_{l-1}=i_{l-1}],
$$
i.e., the values of $X_{l-2},...,X_{1}$ in no way
affect the value of $X_l$, given the value of $X_{l-1}$. 
Such a discrete-time process is
called a Markov chain. It is said to be
~stationary if the probability of going from one state to another is
independent of the time at which the ~transition is being made. That is, for all
states $i$ and $j$,
$$
P[X_l=j\vert X_{l-1}=i]=P[X_{l+k}=j\vert X_{l+k-1}=i]
$$
for $k=-(l-1),-(l-2),...,-1,0,1,2,...$. In this case we set
$p_{ij}\equiv P[X_l=j\vert X_{l-1}=i]$ and call $p_{ij}$ the transition
probability for going from state $i$ to $j$.

For a discrete time stationary Markov chain $\{X_i\}$, $i\in\Nb$, with a finite
state space $S=\{1,2,3,...,m\}$, there are $m^2$ transition
probabilities $\{p_{ij}\}$, $i,j=1,2,...,m$.  $P=(p_{ij})$ is called the
transition matrix corresponding to the discrete-time stationary Markov
chain $\{X_i\}$. The transition matrix $P$ has the following properties:
\be\label{pm}
p_{ij}\ge 0,~~~\sum_{i=1}^m p_{ij}=1,~~~i,j=1,2,...,m.
\ee
Any square matrix that satisfies condition (\ref{pm}) is called
a stochastic matrix.

A continuous-time stochastic process,
$\{X_t\}_{t\in\Rb_+}$ is said to satisfy the Markov property if for all
times $t_0<t_1<...<t_l<t$ and for all $l$ it is true that
$$
P[X_t=j\vert X_{t_0}=i_0,X_{t_1}=i_1,...,X_{t_l}=i_l]
=P[X_t=j\vert X_{t_l}=i_l].
$$
Such a process is called a continuous-time Markov chain. It
is said to be stationary if for
every $i$ and $j$ the transition function, $P[X_{t+h}=j\vert X_{t}=i]$,
is independent of $t$. In this case $P(t)=P(X_{t=j}\vert X_0=i)$ is a
semigroup (e.g. on $l^2(S)$), called transition semigroup associated
with the Markov chain. Its generator $Q=(q_{ij})$ has the properties:
\be\label{qm}
q_{ij}\ge 0,~~~i\neq j,~~~~q_{ii}=-\sum_{i\neq j}q_{ij},~~~i,j=1,2,...,m
\ee
and is called an intensity matrix. Vice versa, any $Q$ 
(satisfying (\ref{qm}) and properly defined as a closed operator when $S$ is
infinite) gives rise to a unique
continuous-time transition semigroup, $P(t)=e^{Qt}$, $t\geq 0$,
which can be interpreted
as transition semigroup associated to a certain Markov chain
(with state space $S$) \cite{markov}.

The properties of Markov chains are determined by the transition matrix
$P$ for discrete-time stochastic process and the intensity matrix $Q$ for
continuous-time stochastic processes.
If the eigenvalues and eigenstates of $P$ and $Q$ are
known, then exact results related to the stochastic processes, such as
time-dependent averages and correlations, can be obtained.

Now we consider a chain (in the algebraic sense of sections 1-3)
with $L+1$ sites. To every site $i$ of the chain we associate $n+1$
states described by the variable $\t_i$ taking $n+1$ integer values,
\be\label{tau}
\t_i\equiv (\t_i^0=0,\t_i^1=1,\t_i^2=2,...,\t_i^n=n),
\ee
(conventionally a vacancy at site $i$ is associated with the state $0$).
We associate to the lattice site $i$ of the algebraic chain a Hilbert space of
dimension $n+1$. The state space of the algebraic chain 
is then finite and has a total of $(n+1)^{L+1}$ states.

By definition, for an integrable chain model with Hamiltonian $H$ one has 
exact solutions for the
eigenvalues and eigenstates of $H$. The system  remains integrable
if one adds to $H$ a constant term $c$ and multiplies $H$ by a constant
factor $c^\p$. Moreover the eigenvalues of $H$
will not be changed if one changes the local basis, i.e., the following
Hamiltonian $H^\p$,
\be\label{hp}
H^\p=BHB^{-1},~~~~B=\displaystyle\otimes_{i=1}^{L+1}B_i,
\ee
where $B_i\equiv b$ and $b$ is an $(n+1)\times (n+1)$ non singular matrix, 
has the same eigenvalues as $H$.
Therefore if an integrable chain model with Hamiltonian $H$ can be
transformed by $B$ (modulo $c$, $c^\p$) 
into a stochastic matrix $P$, in the sense that
\be\label{pmt}
P=B(c^\p H+c\R1)B^{-1},
\ee
where $\R1$ is the $(n+1)^{L+1}\times (n+1)^{L+1}$ identity matrix,
$B$ as in (\ref{hp}), such that $P$ satisfies (\ref{pm}) (with
$m=(n+1)^{L+1}$),
then $P$ defines a discrete-time Markov chain and the related stochastic
process can be simply
studied by using the properties of the related
integrable model with Hamiltonian $H$. 

And if an integrable chain model with Hamiltonian $H$ can be
transformed (modulo $c$, $c^\p$) into an intensity matrix $Q$, in the
sense that
\be\label{qmt}
Q=B(c^\p H+c\R1)B^{-1}
\ee
with $Q$ satisfying (\ref{qm}) (with $m=(n+1)^{L+1}$),
then $Q$ determines a continuous-time Markov chain and its
properties can also be obtained by using the results of the
related integrable Hamiltonian $H$. 

In the following we discuss the question of wether the integrable models 
obtained in the way presented in this paper could
be transformed into stationary Markov chains through
transformations of the forms (\ref{pmt}) or (\ref{qmt}). Of cause a 
stochastic matrix $P$ (i.e. as given by (\ref{pm})) can not in general be
transformed into a intensity matrix $Q$ (as given by (\ref{qm})) by the spectrum preserved
similarity transformation, $P=B(c^\p Q+c\R1)B^{-1}$. That is, an
integrable chain model with Hamiltonian $H$ that gives rise to a
discrete-time Markov chain by the transformation (\ref{pmt}) will in general 
not give rise to a ~continuous time Markov chain by (\ref{qmt}), and vice
versa.

\subsection{Discrete-time Markov Chains Related to $A_n$ Symmetric Integrable 
Models}

We first note that for an integrable chain model with Hamiltonian
$H=\sum_{i=1}^L h_{i,i+1}$ and $(n+1)$ states at every site $i$,
$i=1,2,...,L+1$, if the sum of the elements in any row of the 
$(n+1)^2\times (n+1)^2$ matrix
$h$ is $1/L$, then the sum of the elements in any row of 
the matrix $H$ is $1$. Hence if under the following transformation $h\to
h^\p$ given by
\be
h^\p=(b\otimes b)(c^\p h+c \1\otimes\1) 
(b^{-1}\otimes b^{-1}),
\ee
the sum of the elements in any row of $h^\p$ is $1/L$ and
$(h^\p)_{\alpha,\beta}\geq 0$, $\alpha,\beta=1,2,...,(n+1)^2$, for 
some real constants $c^\p, c$ and a non
singular $(n+1)\times (n+1)$ matrix $b$, then
$P=\sum_{i=1}^L h^\p_{i,i+1}$ defines a stationary discrete-time Markov
chain. $P$ has the same eigenvalue spectrum (shifted by a constant)
as the spectrum of the integrable model with Hamiltonian $H$. If $P$ is
invariant under a certain algebra $A$, we call the Markov chain $A$
symmetric.

{\sf [Theorem 6]}. The following matrix
\be\label{pan}
\ba{rcl}
P_{A_n}&=&\displaystyle\frac{1}{L(n+1)}H_{A_n}
=\displaystyle\frac{1}{L(n+1)}\sum_{i=1}^L
(\Delta C_{A_n}+\1\otimes\1)_{i,i+1}\\[4mm]
&=&\displaystyle\frac{1}{L(n+1)}
\displaystyle\sum_{i=1}^L\left[(n+1)\displaystyle\sum_{\alpha=1}^{n(n+1)/2}
((e_\alpha)_i(f_\alpha)_{i+1}+(f_\alpha)_i(e_\alpha)_{i+1})\right.\\[5mm]
&&+\displaystyle\sum_{\alpha=1}^{n}\alpha(n+1-\alpha)
(h_\alpha)_i(h_\alpha)_{i+1}\\[5mm]
&&\left.+\displaystyle\sum_{\alpha=1}^{n}\displaystyle\sum_{\beta=1}^{n-\alpha}
\alpha(n+1-\alpha-\beta)((h_\alpha)_i(h_{\alpha+\beta})_{i+1}
+(h_{\alpha+\beta})_i(h_{\alpha})_{i+1})\right]+\displaystyle\frac{1}{(n+1)}
\ea
\ee
defines a stationary discrete-time $A_n$ symmetric Markov chain.

{\sf [Proof]}. I. Set $h^\p\equiv\frac{1}{L(n+1)}(\Delta
C_{A_n}+\1\otimes\1)$. Then 
\be\label{pform}
P_{A_n}=\sum_{i=1}^L h^\p_{i,i+1}.
\ee 
From formula (\ref{48}) we have
\be\label{lhp}
\ba{rcl}
(h^\p)_{\alpha\beta}&=&\displaystyle\frac{1}{L(n+1)}(\Delta
C_{A_n}+\1\otimes\1)_{\alpha\beta}\\[4mm]
&=&\displaystyle\frac{1}{L(n+1)}
[\delta_{\alpha\beta}[(n+1)\delta_{\alpha,l(n+1)+l+1}-1]\\[4mm]
&&+(n+1)[\delta_{\alpha,j(n+2)+k+2}\delta_{\beta,(j+1)(n+2)+k(n+1)}\\[4mm]
&&+\delta_{\beta,j(n+2)+k+2}\delta_{\alpha,(j+1)(n+2)+k(n+1)}]+\delta_{\alpha\beta}]\\[4mm]
&=&\displaystyle\frac{1}{L}
[\delta_{\alpha\beta}\delta_{\alpha,l(n+1)+l+1}
+\delta_{\alpha,j(n+2)+k+2}\delta_{\beta,(j+1)(n+2)+k(n+1)}\\[4mm]
&&+\delta_{\beta,j(n+2)+k+2}\delta_{\alpha,(j+1)(n+2)+k(n+1)}]\geq 0.
\ea
\ee
Therefore $(P_{A_n})_{\alpha\beta}\geq 0$,
$\alpha,\beta=1,2,...,(n+1)^2$.

II. By using the identity (\ref{identity}), we get
$$
\ba{rcl}
\displaystyle\sum_{\beta=1}^{(n+1)^2}(h^\p)_{\alpha\beta}
&=&\displaystyle\frac{1}{L(n+1)}\displaystyle\sum_{\beta=1}^{(n+1)^2}
[(n+1)\delta_{\alpha\beta}\delta_{\alpha,l(n+1)+l+1}\\[5mm]
&&+(n+1)[\delta_{\alpha,j(n+2)+k+2}\delta_{\beta,(j+1)(n+2)+k(n+1)}\\[4mm]
&&+\delta_{\beta,j(n+2)+k+2}\delta_{\alpha,(j+1)(n+2)+k(n+1)}]]\\[4mm]
&=&\displaystyle\frac{1}{L(n+1)}(n+1)=\displaystyle\frac{1}{L}.
\ea
$$
Hence the sum of the elements of any row of the matrix $P_{A_n}$ is one,
i.e., $\sum_{\beta=1}^{(n+1)^{L+1}}(P_{A_n})_{\alpha\beta}
=\sum_{\beta=1}^{(n+1)^{L+1}}(\sum_{i=1}^L h^\p_{i,i+1})_{\alpha\beta}=1$.

III. As $H_{A_n}$ is invariant under $A_n$,
$P_{A_n}=\frac{H_{A_n}}{L(n+1)}$ is obviously invariant under $A_n$
and has the same spectrum as $H_{A_n}$.

By the definition (\ref{pm}) $P_{A_n}$ is the transition matrix of a
stationary discrete-time $A_n$ symmetric Markov chain. 
\hfill $\rule{3mm}{3mm}$

We give some discussions on the properties of the stationary discrete-time 
$A_n$ symmetric Markov chain associated with the 
stochastic matrix $P_{A_n}$. The state
space of this Markov chain is,
$S=(1,2,...,(n+1)^{L+1})$, which corresponds to $(n+1)^{L+1}$
states, 
\be\label{taut}
(\t_0\otimes \t_1\otimes...\otimes \t_n),
\ee
$\t_i$ as in (\ref{tau}), of the algebraic chain with $L+1$ lattice sites. 

The properties of a Markov chain are determined
by the transition matrix $P=(p_{ij})$. A subset $C$ of 
the state space $S$ is called closed if $p_{ij}=0$ for all $i\in C$ and
$j\not\in C$. If a closed set consists of a single state, then that
state is called an absorbing state. A Markov chain is called irreducible
if there exists no nonempty closed set other than $S$ itself.

From formula (\ref{lhp}) we have 
\be\label{hpp}
\ba{l}
(h^\p)_{\alpha\alpha}=(h^\p)_{(n+1)^2,(n+1)^2}=\frac{1}{L},~~~
(h^\p)_{\alpha\beta}=(h^\p)_{\beta\alpha}=0,~~~\beta\neq\alpha,\\[3mm]
\alpha=l(n+1)+l+1,~~~l=0,1,...,n.
\ea
\ee

Let 
\be\label{s0}
S_0=\left(\alpha\vert\alpha=l\frac{(n+1)((n+1)^L-
1)+n}{n}+1\right),~~~l=0,1,...,n,
\ee
be a subset of the state space $S$.
From formula (\ref{pform}), with 
$$
(h^\p)_{i,i+1}=\1_1\otimes\1_2\otimes...\otimes\1_{i-1}\otimes
h^\p\otimes\1_{i+2}\otimes...\otimes\1_{L+1},
$$
we get
\be\label{panp}
(P_{A_n})_{\alpha\alpha}=1,~~~
(P_{A_n})_{\beta\alpha}=(P_{A_n})_{\alpha\beta}=0,~~~\beta\neq\alpha,
~~~\alpha\in S_0.
\ee

Therefore the $n+1$ states in $S_0$ are absorbing states of the
Markov chain $P_{A_n}$. This chain is by definition reducible. For a
reducible Markov chain the ``long time" probability distribution, if it
exists, may depend on the initial conditions, i.e.,
$\lim_{l\to\infty}(P_{A_n})^{l}_{\gamma\beta}$ may depend on $\gamma$.
From the properties (\ref{panp}) of $P_{A_n}$, we see that if the Markov
chain $P_{A_n}$ is initially at one of the states $\alpha\in S_0$, it
will remain in that state $\alpha$ forever. These $n+1$ absorbing states 
correspond to the states of the algebraic chain through (\ref{taut}).
For instance, the states $1$ and $(n+1)^{L+1}$ in $S$ correspond to the
states $(0,0,...,0)$ (all the
sites of the algebraic chain are at state $0$) and $(n,n,...,n)$ (all the
sites of the algebraic chain are at state $n$).

\subsection{Continuous-time Markov Chains Related to $A_n$ Symmetric Integrable 
Models}

For an integrable chain model with Hamiltonian 
$H=\sum_{i=1}^L h_{i,i+1}$ and with $(n+1)$ states at every site of the chain, 
if the sum of the elements in any column of the matrix
$h$ is $0$, then the sum of the elements in any column of 
the matrix $H$ is also $0$. Hence if under the following transformation
$h\to h^{\p\p}$ with:
\be
h^\pp=(b\otimes b)(c^\p h+c \1\otimes\1)(b^{-1}\otimes b^{-1}),
\ee
the sum of the elements in any column of $h^\pp$ is $0$ and
$(h^\pp)_{\alpha,\beta}\geq 0$, $\alpha\neq\beta=1,2,...,(n+1)^2$, for
some real constants $c^\p, c$ and a non
singular $(n+1)\times (n+1)$ matrix $b$, then
$Q=\sum_{i=1}^L h^\pp_{i,i+1}$ is the intensity matrix for some stationary 
continuous-time Markov
chain. $Q$ has the same eigenvalue spectrum (shifted by a constant)
as the spectrum of the Hamiltonian $H$. We call the Markov chain $A$ 
symmetric if $Q$ is invariant under the algebra $A$.

{\sf [Theorem 7]}. The following matrix $Q$ is the intensity matrix of a
stationary continuous-time Markov chain,
\be\label{qan}
\ba{rcl}
Q_{A_n}&=&H_{A_n}-(n+1)L
=\displaystyle\sum_{i=1}^L(\Delta C_{A_n}-n\1\otimes\1)_{i,i+1}\\[4mm]
&=&\displaystyle\sum_{i=1}^L\left[(n+1)\displaystyle\sum_{\alpha=1}^{n(n+1)/2}
((e_\alpha)_i(f_\alpha)_{i+1}+(f_\alpha)_i(e_\alpha)_{i+1})
+\displaystyle\sum_{\alpha=1}^{n}\alpha(n+1-\alpha)
(h_\alpha)_i(h_\alpha)_{i+1}\right.\\[5mm]
&&\left.+\displaystyle\sum_{\alpha=1}^{n}\displaystyle\sum_{\beta=1}^{n-\alpha}
\alpha(n+1-\alpha-\beta)((h_\alpha)_i(h_{\alpha+\beta})_{i+1}
+(h_{\alpha+\beta})_i(h_{\alpha})_{i+1})\right]-nL.
\ea
\ee

{\sf [Proof]}. Set $h^\pp=(\Delta C_{A_n}-n\1\otimes\1)$. Then 
\be\label{qanhpp}
Q_{A_n}=\sum_{i=1}^L h^\pp_{i,i+1}.
\ee
From (\ref{48}) we observe that, for $\alpha\neq\beta$,
$$
\ba{ll}
h^\pp_{\alpha\neq\beta}=
&(n+1)[\delta_{\alpha,j(n+2)+k+2}\delta_{\beta,(j+1)(n+2)+k(n+1)}\\[3mm]
&+\delta_{\beta,j(n+2)+k+2}\delta_{\alpha,(j+1)(n+2)+k(n+1)}]\geq 0.
\ea
$$
Therefore $(Q_{A_n})_{\alpha\neq\beta}\geq 0$, $\alpha,\beta=1,2,...,
(n+1)^{L+1}$.

Again by (\ref{48}) the sum of the elements in any given column $\beta$
of the matrix $h^\pp$ is
$$
\ba{l}
\displaystyle\sum_{\alpha=1}^{(n+1)^2}
h^\pp_{\alpha\beta}=\displaystyle\sum_{\alpha=1}^{(n+1)^2}(n+1)
[\delta_{\alpha\beta}(\delta_{\alpha,l(n+1)+l+1}-1)\\[5mm]
~~~+(n+1)[\delta_{\alpha,j(n+2)+k+2}\delta_{\beta,(j+1)(n+2)+k(n+1)}
+\delta_{\beta,j(n+2)+k+2}\delta_{\alpha,(j+1)(n+2)+k(n+1)}]\\[3mm]
=\displaystyle\sum_{\alpha\neq l(n+1)+l+1}^{(n+1)^2}(n+1)
[-\delta_{\alpha\beta}
+\delta_{\alpha,j(n+2)+k+2}\delta_{\beta,(j+1)(n+2)+k(n+1)}\\[5mm]
~~~+\delta_{\beta,j(n+2)+k+2}\delta_{\alpha,(j+1)(n+2)+k(n+1)}]\\[3mm]
=(n+1)(-1+\delta_{\beta,(j+1)(n+2)+k(n+1)}\vert_{\alpha=j(n+2)+k+2}
+\delta_{\beta,j(n+2)+k+2}\vert_{\alpha=(j+1)(n+2)+k(n+1)})\\[3mm]
=0,
\ea
$$
i.e., the sum of the elements in any given column $\beta$,
$\beta=1,2,...,(n+1)^2$, of the matrix $h^\pp$ is zero.
~Therefore the sum of the elements in any given column $\beta$,
$\beta=1,2,...,(n+1)^{L+1}$, of the matrix $Q_{A_n}$ is also zero,
$\sum_{\alpha=1}^{(n+1)^{L+1}}(Q_{A_n})_{\alpha\beta}=0$.
At last $Q_{A_n}=H_{A_n}-(n+1)L$ is obviously $A_n$ symmetric with the
same spectrum (shifted by a constant) as $H_{A_n}$. \hfill $\rule{3mm}{3mm}$

The long run distribution of the Markov chain described in Theorem 7
is given by the vector ${\bf \pi}=(\pi_1,\pi_2,...)$, where
$\pi_i$ represents the ``long time" probability of the state $i\in S$, 
satisfying
\be\label{pi}
\sum_{\alpha=1}^{(n+1)^{L+1}} (Q_{A_n})_{\alpha\beta}\pi_\alpha=0,
~~~\forall\beta\in S,~~~\sum_{\alpha=1}^{(n+1)^{L+1}}\pi_\alpha=1.
\ee
However as this Markov chain is not irreducible, the solution of the
equation (\ref{pi}) is not unique but depends on the initial conditions.
From (\ref{48}) we see that
$$
\ba{l}
(h^\pp)_{\alpha\beta}=(h^\pp)_{\beta\alpha}=0,~~~\forall \beta,\\[3mm]
\alpha=l(n+1)+l+1,~~~l=0,1,...,n.
\ea
$$
Hence from (\ref{qanhpp}) we get
$$
(Q_{A_n})_{\alpha\beta}=(Q_{A_n})_{\beta\alpha}=0,~~~\alpha\in S_0,~~
\forall \beta,
$$
with $S_0$ as in (\ref{s0}).
Therefore if this Markov chain is initially at a given state $\alpha\in
S_0$, it will remain at that state. 

The states $\beta\not\in S_0$ form a closed subset of $S$.
From (\ref{48}) and (\ref{qanhpp}) one also learns that the absolute value
of all the nonzero elements of any column of the intensity 
matrix $Q_{A_n}$ are equal. Let $S^\p$ be a closed subset of $S$ 
with $l$ elements. If the Markov chain is initially in the closed set
$S^\p$, then it will remain in $S^\p$ and the long run distribution
is ${\bf \pi}=(\pi_1,\pi_2,...,\pi_{(n+1)^{L+1}})$, where $\pi_i=1/l$
for $i\in S^\p$ and $\pi_i=0$ if $i\not\in S^\p$.

\section{Conclusion and Remark}

Using the Casimir operators and coproduct operations of algebras, we have
given a simple way to construct chain models with a certain 
algebraic symmetry and nearest or
non-nearest neighbours interactions. We discussed integrable 
chain models with nearest neighbours interactions with the
symmetries provided by the fundamental representation of the classical 
Lie algebra $A_n$. 
It is shown that corresponding to these $A_n$ symmetric integrable chain 
models there are exactly solvable stationary discrete-time (resp.
continuous-time) Markov chains whose spectra of the
transition matrices (resp. intensity matrices)
are the same as the ones of the corresponding integrable models.

Other symmetric integrable models ( e.g. with $B_n$,
$C_n$, $D_n$ symmetry) and related TL algebraic structures
and Markov chains can be investigated in a similar way. The
discussion of integrable models related to higher
dimensional representations of the algebras and the integrability of
chain models with non-nearest neighbours interactions is postponed to 
further work.

\vspace{2.5ex}
ACKNOWLEDGEMENTS: We would like to thank the A.v. Humboldt
Foundation for the financial support given to the second named author.

\end{document}